\documentclass{jfm}
\usepackage{subcaption}
\usepackage{graphicx}
\usepackage{epstopdf, epsfig}
\usepackage{color,soul}
\usepackage{hyperref}
\usepackage{amsmath}
\usepackage{newtxtext}
\usepackage{newtxmath}

\DeclareMathOperator{\acosh}{arcosh}

\newcommand{\figref}[2][{}]{\hyperref[#2]{\figurename~\ref{#2}#1}} 
\renewcommand{\vec}{\boldsymbol}
\graphicspath{{figures/}}

\shorttitle{sedimenting squirmer near a wall}
\shortauthor{Shum, Palaniappan and Young}

\title{Hydrodynamic interactions between a sedimenting squirmer and a planar wall}

\author{Henry Shum\aff{1},
  D. Palaniappan\aff{2}
 \and Y.-N. Young\aff{3} \corresp{\email{yyoung@njit.edu}}}

\affiliation{\aff{1}Department of Mathematics, University of Waterloo, Waterloo, ON  N2J 3G1, Canada
\aff{2}Department of Mathematics and Statistics, Texas A\&M University - Corpus Christi, TX 78412, USA
\aff{3}Department of Mathematical Sciences, New Jersey Institute of Technology, Newark, NJ, 07102, USA}

\begin{document}

\maketitle

\begin{abstract}
The hydrodynamic interactions between a sedimenting microswimmer and a solid wall have ubiquitous biological and technological applications. A plethora of gravity-induced swimming dynamics near a planar no-slip wall provides a platform for designing artificial microswimmers that can generate directed propulsion through their translation-rotation coupling near a wall.  In this work we provide exact solutions for a squirmer (a model swimmer of spherical shape with a prescribed slip velocity) facing either towards or away from a planar wall perpendicular to gravity. These exact solutions are used to validate a numerical code based on the boundary integral method with an adaptive mesh for distances from the wall down to 0.1\% of the squirmer radius. This boundary integral code is then used to investigate the rich gravity-induced dynamics near a wall,
%with and without repulsion, 
mapping out the detailed bifurcation structures of the swimming dynamics in terms of orientation and distance to the wall. Simulation results show that a squirmer may transverse along the wall, move to a fixed point at a given height with a fixed orientation in a monotonic way or in an oscillatory fashion, or oscillate in a limit cycle in the presence of wall repulsion. 
% {\bf APS-DFD 2023:}
% Active fluids exhibit complex behaviors that result from combinations of hydrodynamic, chemical, and biological effects. We explore interactions among active particles and between active particles and a solid wall through experiments and numerical simulations. In experiments, droplets of the liquid crystal 4-cyano-4'-pentylbiphenyl (5CB) are added to a concentrated surfactant solution. Solubilization of the droplets leads to self-propelling behavior and observable fluid flows around the droplets. Notably, we find that the droplets, which sediment to the bottom surface under gravity, organize into lattice-like structures. As a model for active droplets and swimming organisms, we consider neutral, extensile, and contractile spherical squirmers placed near a no-slip plane boundary and compute their trajectories in Stokes flow using boundary element method simulations. We describe transitions in their individual and collective behavior as their excess mass density is varied and compare the numerical results with experimental observations.
\end{abstract}

%\begin{keywords}
%Authors should not enter keywords on the manuscript, as these must be chosen by the author during the online submission process and will then be added during the typesetting process (see http://journals.cambridge.org/data/\linebreak[3]relatedlink/jfm-\linebreak[3]keywords.pdf for the full list)
%\end{keywords}

\section{Introduction}
 Microswimmers behave very differently near a wall as their interactions with a solid boundary alter their speed, direction, and how they interact with each other 
\citep{shum2010modelling,takagi2014hydrodynamic,elgeti2016microswimmers}, giving rise to many interesting phenomena, such as the swirling of bacteria next to a substrate and clustering of phoretic Janus particles and bacteria near a wall. While the far-field flow attracts and aligns a pusher (puller) to move along (normal to) a no-slip wall, near-field hydrodynamics, steric interactions and contact dynamics give rise to wall scattering with the swimmer escaping from the wall at a characteristic angle that is independent of the initial direction of approach to the wall \citep{berke2008hydrodynamic,li2009accumulation}.

The hydrodynamic interactions between a microswimmer and a solid wall are more complex when the swimmer sediments to the wall under gravity (due to the density mismatch between the swimmer and the surrounding fluid). Several types of dynamics of a sedimenting swimmer have been reported: scattering (escaping) from the wall, swimming along the wall at a fixed distance and tilted orientation, and periodic bouncing on the wall \citep{or2009dynamics,crowdy2010two}. Under gravity, artificial surface walkers or micro rollers stay close to the wall. At the same time, they rotate under an external force field, exploiting their interactions with a solid surface to generate directed propulsion \citep{tierno2008controlled,sing2010controlled,driscoll2017unstable}. These microswimmers are easy to manipulate for directed transport and offer wide applications in targeted therapeutics and microsurgery \citep{alapan2020multifunctional,ahmed2021bioinspired}.

The interactions between a flagellated swimmer and a flat solid wall have been modeled using a multipole approach \citep{spagnolie2012hydrodynamics}, which is shown to give good agreement with boundary integral simulations. Through the contribution of each singularity to the effect of a wall on the flagellated swimmer,  the reduced model captures the main dynamic features of the wall-induced hydrodynamics of a microswimmer. 
Alternatively, a microswimmer is often simplified and modeled as a squirming sphere with a slip velocity on the surface to mimic the surrounding flow created by the layer of beating cilia on the microswimmer~\citep{lighthill_squirming_1952, blake_spherical_1971}. Such simplification allows the usage of the Lorentz reciprocal theorem to derive an exact solution for a squirming sphere close to a no-slip surface \citep{papavassiliou2017exact}.  Exact solutions for a sphere moving towards or away from a flat wall have been derived by \cite{brenner1961slow}. \cite{cox1967slow} later derived the near-field solution that shows the asymptotic  divergence of the viscous drag coefficient as the sphere approaches the solid wall. For a squirmer interacting with a solid wall through the hydrodynamic interactions, \cite{thery2023hydrodynamic} combined the far-field flow of a sedimenting sphere \citep{kim2013microhydrodynamics} with the squirming flow to illustrate the different swimming dynamics near a wall and noted that the far-field approximation may not be uniformly valid across different types of squirmer dynamics. For example, oscillatory (bouncing) dynamics of a squirmer may involve motion both near to and far from the wall, and the sliding squirmer can also occur in the near-field \citep{li2014hydrodynamic,ruhle2018gravity,kuhr2019collective}. 

In this work we seek to elucidate the detailed swimming dynamics of a single squirmer sedimenting toward a flat wall, using a boundary integral code validated for both the near-field and far-field hydrodynamics. In particular, we seek to quantify how the swimming dynamics of a squirmer sedimenting to a wall depends on $\alpha$ (the ratio of sedimenting velocity to swimming speed) and $\beta$ (the ratio of the first two squirming mode amplitudes). We first derive exact solution for a squirming sphere sedimenting towards a flat no-slip wall, using the approach in \cite{brenner1961slow}. We use this analytic solution to validate a boundary integral code, which is highly efficient and accurate for us to examine the dynamics of a sedimenting squirmer near a flat wall over a wide range of parameters.

This paper is organized as follows. In \S~\ref{sec:Problem_Formulation} we present the formulation for a squirmer under gravity in the presence of a planar bottom wall. We assume the squirmer is immersed in a viscous Stokes flow, and there may be a steric repulsion between the solid wall and the squirmer when close to the wall. We summarize the boundary integral formulation for the numerical implementation in \S~\ref{subsec:numerical_algorithm}. We present the exact solution for a squirmer perpendicular to a flat no-slip wall in \S~\ref{sec:squirmer_wall}. This exact solution allows us to derive an extended far-field formula, which we compare against the numerical results to validate the boundary integral code, and also examine the range of validity for the far-field approximation in \S~\ref{subsec:far-field}. We further study the near-field approximation to the flow and compare between the exact solution, boundary integral simulation results, and the asymptotic results in the literature in \S~\ref{subsec:near-field}. In \S~\ref{sec:general_swimming_dynamics} we classify the swimming dynamics of a squirmer interacting with a no-slip planar wall under gravity, and show the detailed bifurcation structures for mixed squirming modes (\S~\ref{subsec:mixed_modes}) and pure squirming modes (\S~\ref{subsec:pure_modes}). In \S~\ref{sec:conclusion} we provide discussions of our results and implications for future directions.

\section{Problem Formulation}
\label{sec:Problem_Formulation}
\begin{figure}
    \centering
    \includegraphics[width=0.5\textwidth]{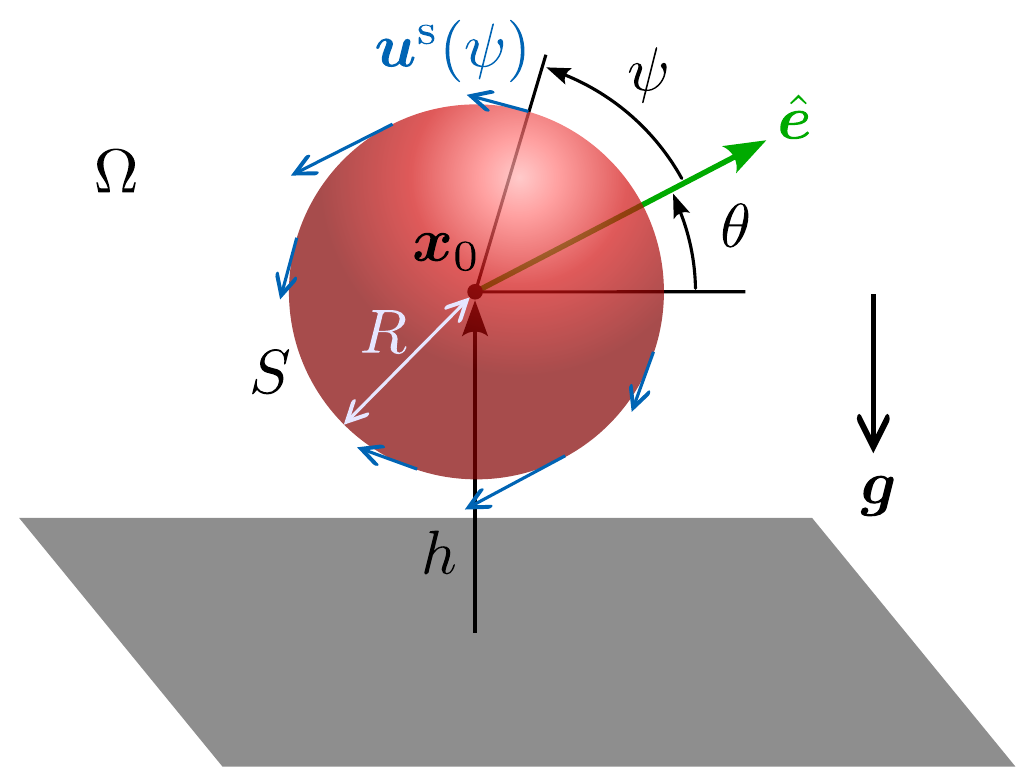}
    \caption{Schematic of a spherical squirmer of radius $R$ at a height $h$ above a no-slip planar boundary with gravity pointing toward the wall.}
    \label{Fig1}
\end{figure}

We consider a three-dimensional incompressible viscous fluid governed by the equations of Stokes flow,
\begin{equation}
\label{eq:StokesFlow}
    -\nabla p + \mu_f\nabla^2\vec{u} = \vec{0}, \quad \nabla \cdot\vec{u} = 0,
\end{equation}
for ${\bf x}\in\Omega$, the space between squirmers and a planar wall in Fig.~(\ref{Fig1}). The squirmer is a sphere of radius $R$ located at a height $h$ above the planar wall, and has an orientation vector $\hat{\vec{e}}$ at an angle $\theta$ with respect to the wall: $\theta=0$ when the squirmer is parallel to the wall, and $\theta=\pi/2$ when the squirmer is upright. Furthermore we assume that there is a density mismatch $\Delta \rho$ between the viscous fluid and the squirmer, which sediments under gravity. For spherical squirmers of constant excess density $\Delta \rho$ relative to the surrounding fluid, the excess gravitational force on the squirmer is given by
\begin{equation}
	\vec{F}^\mathrm{grav} = -F^\mathrm{grav}\, \hat{\vec{e}}_z = -\frac{4}{3}\pi R^3 g \Delta\rho\, \hat{\vec{e}}_z,
\end{equation} 
where $g$ is the constant gravitational acceleration in the $-\hat{\vec{e}}_z$ direction, see Fig.~(\ref{Fig1}).

We focus on fluid flow generated by the activity on the squirmer surface $S$, and assume that the fluid flow vanishes in the far-field. Furthermore we consider spherical squirmer with up to the first two squirming modes, prescribing either the tangential velocity or the tangential stress distribution. The prescribed surface tangential velocity distribution is purely in the polar ($\hat{e}_\psi$) direction and is given by
\begin{equation}
\label{eq:slip_velocity}
	\vec{u}^\mathrm{s} = u_\psi \hat{\vec{e}}_\psi, \quad u_\psi = B_1 \sin(\psi) + B_2 \sin(\psi)\cos(\psi),
\end{equation}
where $\psi$ is the angle of the radial vector at a point on the surface to the orientation vector $\hat{\vec{e}}$ of the squirmer, see Fig.~(\ref{Fig1}). $B_1$ is the neutral, self-propelling mode with swimming speed $V = 2B_1/3$ in free space ($V=2/3$ when $B_1 = 1$). $B_2$ is the stresslet mode: $B_2 > (<) 0$ for contractile puller (extensile pusher) squirmers.  The velocity continuity and stress balance at the squirmer boundary and the planar wall provide the boundary conditions that close the system of equations.

%\subsection{Short-range repulsion}\label{subsec:repulsion}
In the simulations, we apply a repulsive force on the squirmer at a distance $r$ (the bottom of the sphere to the wall) to the no-slip wall \citep{brady1985rheology,ishikawa2006hydrodynamic}
\begin{equation}
\label{eq:repulsion}
	F^\mathrm{rep} = \frac{C^\mathrm{rep} \exp(-a^\mathrm{rep} r)}{1-\exp(- a^\mathrm{rep} r)},
\end{equation}
pointing away from the wall. We use numerical values $C^\mathrm{rep} = 10^3, a^\mathrm{rep} = 100$ to ensure a short-range repulsion, sufficient to maintain an equilibrium separation of $0.04R$ for a sphere of radius $R=1$ and free space sedimentation speed $V_g=1$ with no active squirming.

%%%%%%%%%%%%%%%%%%%%%%%%%%%%%%%%%%%%%%%%%%%%%%%%%%%%%%%%%%%%%%%%%%%%%%%
%\subsection{Gravity}

\subsection{Numerical algorithm and validation} 
\label{subsec:numerical_algorithm}
% We consider three-dimensional fluid flow governed by the equations of incompressible Stokes flow,
% \begin{equation}
%     -\nabla p + \mu_f\nabla^2\vec{u} = \vec{0}, \quad \nabla \cdot\vec{u} = 0.
% \end{equation}
The incompressible velocity field $\vec{u}$ that satisfies the Stokes equations (Eqs.~(\ref{eq:StokesFlow})) can be expressed in terms of integrals of force and/or stress densities on the surfaces~\citep{pozrikidis1992boundary}. In particular, 
in the absence of a background flow, the $i$th component of the fluid velocity at a point $\vec{x}$ outside or on the surface $S$ of a particle can be represented by a single-layer potential as
\begin{equation}
    u_i(\vec{x}) = \int_S \mathbf{G}_{ij}(\vec{x},\vec{y})q_j(\vec{y})\,\mathrm{d}S(\vec{y}),
    \label{eq:slp}
\end{equation}
where $\mathbf{G}_{ij}$ are the $ij$th components of the Green's function tensor $\mathbf{G}$ for Stokes flow. In free space, the Green's function has the formula
\begin{equation}
    \mathbf{G}^\mathrm{FS}_{ij}(\vec{x},\vec{y}) = \frac{\delta_{ij}}{r} + \frac{r_i r_j}{r^3},
\end{equation}
where $\vec{r} = \vec{x}-\vec{y}$ and $r = |{\vec{r}}|$. For simulations near a no-slip plane boundary, we use the modified Green's function $\mathbf{G} =\mathbf{G}^\mathrm{FS}+\mathbf{G}^\mathrm{im}$ so that the velocity field satisfies the no-slip boundary condition on the plane $z=0$ by including image terms given by~\citep{blake_note_1971}
\begin{equation}
    \mathbf{G}^\mathrm{im}_{ij}(\vec{x},\vec{y}) = -\frac{\delta_{ij}}{\tilde{r}} - \frac{\tilde{r}_i \tilde{r}_j}{\tilde{r}^3} + 2y_3\left(\delta_{j \alpha} \delta_{\alpha l}-\delta_{j 3}\delta_{3 l}\right)\frac{\partial}{\partial \tilde{r}_l}\left[\frac{y_3 \tilde{r}_i}{\tilde{r}^3}-\left(\frac{\delta_{i3}}{\tilde{r}} + \frac{\tilde{r}_i\tilde{r}_3}{\tilde{r}^3}\right)\right],
\end{equation}
where $\tilde{\vec{r}} = (x_1-y_1,x_2-y_2,x_3+y_3)$, $\tilde{r}=|\tilde{\vec{r}}|$, and summations are implied over $\alpha = 1,2$, and $l=1,2,3$.
In the case of a rigid body motion of the particle, the density $\vec{q}$ of the single-layer potential is proportional to the traction vector $\vec{f}$,
\begin{equation}
    \vec{q} = -\frac{1}{8\pi\mu_f}\vec{f}.
\end{equation}
%For a droplet with the same viscosity as the surrounding fluid, the density $\vec{q}$ can be interpreted as the traction jump across the interface,
%\begin{equation}
%    \vec{q} = -\frac{1}{8\pi\mu_f}\Delta \vec{f} = -\frac{1}{8\pi\mu_f}\left(\vec{f}^{\mathrm{out}} - \vec{f}^{\mathrm{in}}\right).
%\end{equation}

For a squirmer with tangential surface velocity distribution $\vec{u}^\mathrm{s}$ moving with translational velocity $\vec{U}$ and rotational velocity $\vec{\Omega}$ about its center $\vec{x}_0$, we have the boundary condition
\begin{equation}
    \vec{U} + \vec{\Omega}\times(\vec{x}-\vec{x}_0) + \vec{u}^\mathrm{s} = \int_S \mathbf{G}(\vec{x},\vec{y})\vec{q}(\vec{y})\,\mathrm{d}S(\vec{y})
    \label{eq:boundary_velocity}
\end{equation}
for $\vec{x}\in S$.

The total hydrodynamic force acting on the particle or droplet is given by
\begin{equation}
    \vec{F}^\mathrm{hydro} = -{8\pi\mu_f}\int_S \vec{q}(\vec{y}) \,\mathrm{d}S(\vec{y})
\end{equation}
and the total hydrodynamic torque is
\begin{equation}
    \vec{L}^\mathrm{hydro} = -{8\pi\mu_f}\int_S \vec{y}\times \vec{q}(\vec{y}) \,\mathrm{d}S(\vec{y}).
\end{equation}

For squirmers that experience forces due to gravity and short-range repulsion, we impose the force balance equation
\begin{equation}
    \vec{F}^\mathrm{hydro} + \vec{F}^\mathrm{rep} + \vec{F}^\mathrm{grav} = \vec{0}.
    \label{eq:force_balance}
\end{equation}
By symmetry, gravity and short-range repulsion do not exert torques on the spherical squirmers so the torque balance equation is
\begin{equation}
\vec{L}^\mathrm{hydro} = \vec{0}.
\label{eq:torque_balance}
\end{equation}

The full system of equations to be solved at a given time consists of \eqref{eq:boundary_velocity}, \eqref{eq:force_balance}, and \eqref{eq:torque_balance}. To solve this system numerically, the surface of the squirmer is discretized into quadratic triangular elements and the density $\vec{q}$ is approximated by a quadratic interpolation of the values at the six nodes of each triangular element. In this work we prescribe the tangential velocities $\vec{u}^s$ in \eqref{eq:boundary_velocity} at each of the $N$ nodes on the surface of the squirmer,  and the single-layer density $\vec{q}$ at the nodes are unknowns, as are the translational and rotational velocity vectors. This yields $3N$ equations in $(3N + 6)$ unknowns. Six further equations arise from the force and torque balance constraints.

\section{Exact solutions for a squirmer perpendicular to a planar wall under gravity}
\label{sec:squirmer_wall}
We next assume that the squirmer orientation vector $\hat{\vec{e}}$ is normal to the planar wall, pointing towards the wall ($\theta=-\pi/2$). Under such axial symmetry, we compute the axisymmetric flow generated by the squirmer interacting with a planar wall using bipolar spherical coordinates defined as
\begin{align}
    z+i\varrho &= i c \cot\frac{1}{2}\left(\eta + i\xi\right),
\end{align}
where $\varrho$ and $z$ are the cylindrical coordinates that can be expressed explicitly in terms of $\eta$ and $\xi$: 
\begin{align}
    \varrho = \frac{c \sin\eta}{\cosh\xi-\cos\eta}, \;\;\; x = \frac{c \sinh \xi}{\cosh\xi - \cos\eta},
\end{align}
with $2c$ the distance between the two poles of the bispherical coordinates.
In the present application it is only necessary to consider the situation $\varrho>0$ which corresponds to $0\leq \xi<\infty$ and $0\le\eta\le\pi$. $\xi=0$ corresponds to a plane, $\xi=\zeta = \acosh(h/R)$ corresponds to a spherical surface of radius $c/\sinh\zeta$ centered at $c\coth\xi$.
In axisymmetric flow, one can write the velocity as a curl of a vector,
${\bf u} = \nabla\times\left(\varphi(\xi,\eta) \hat{e}_{\phi}\right)$, with $\hat{e}_{\phi}$ being the unit vector in the azimuthal direction and $\varphi(\xi,\eta)$ being the 
Stokes stream function. Accordingly, the governing differential equation for the axisymmetric viscous flow
reduces to a fourth order linear differential equation for $\varphi$:
\begin{align}
    D^4\varphi &= (D^2)^2\varphi = 0,
\end{align}
where the differential operator $D^2$ in bispherical coordinates is given by
\begin{align}
    D^2 \equiv \frac{\cosh\xi - \mu}{c^2}\left\{ \frac{\partial}{\partial \xi}\left[\left(\cosh \xi-\mu\right)\frac{\partial}{\partial\xi}\right] + (1-\mu^2)\frac{\partial}{\partial \mu}\left[\left(\cosh \xi-\mu\right)\frac{\partial}{\partial\mu}\right]\right\}
\end{align}
with $\mu\left(\eta\right) = \cos\eta$. The general expression for the stream function $\varphi$ can be given in the bi-spherical coordinates \citep{stimson1926motion}
\begin{align}
    \varphi(\xi,\eta) &= \left(\cosh\xi-\cos\eta\right)^{-3/2}\sum^{\infty}_{n=1}\chi_n\left(\xi\right)V_n\left(\mu\left(\eta\right)\right),\\
    \chi_n\left(\xi\right) &= A_n\cosh\left(n-\frac{1}{2}\right)\xi + B_n\sinh\left(n-\frac{1}{2}\right)\xi + C_n \cosh\left(n+\frac{3}{2}\right)\xi + D_n\sinh\left(n+\frac{3}{2}\right)\xi,\\
    V_n(\mu\left(\eta\right)) &= P_{n-1}(\cos\eta) - P_{n+1}(\cos\eta),
\end{align}
where $P_n$ is the $n$th order Legendre polynomial.
We can express the components of velocity in terms of the stream function as
\begin{align}
    u_{\xi} &= \frac{\left(\cosh\xi-\cos\eta\right)^2}{c\sin\eta}\frac{\partial\varphi}{\partial\eta}\\
    &=-\frac{3}{2c\sqrt{\cosh\xi-\cos\eta}}\sum^{\infty}_{n=1}\chi_n\left(\xi\right)V_n\left(\mu\right) + \frac{\sqrt{\cosh\xi-\cos\mu}}{c \sin\eta}\sum^{\infty}_{n=1}\chi_n\left(\xi\right)\frac{\partial V_n(\mu)}{\partial\eta}, \nonumber \\
    u_{\eta} &= -\frac{\left(\cosh\xi-\cos\eta\right)^2}{c\sin\eta}\frac{\partial\varphi}{\partial \xi} \\
    &=\frac{3\sinh\xi}{2c \sin\eta\sqrt{\cosh\xi-\cos\eta}}\sum^{\infty}_{n=1}\chi_n\left(\xi\right) V_n\left(\mu\right) - \frac{\sqrt{\cosh\xi-\cos\mu}}{c\sin\eta}\sum^{\infty}_{n=1}\chi'_n\left(\xi\right)V_n\left(\mu\right). \nonumber
\end{align}
For each ($n$th) term of the stream function expansion $\varphi$ there are four coefficients $A_n$, $B_n$, $C_n$ and $D_n$ to be determined as a function of the squirmer speed $U$ according to the boundary conditions of the rigid body motion and velocity continuity on the squirmer surface and the planar wall. Assuming axisymmetry (see Fig.~(\ref{Fig1})), we use the force-free condition to compute the squirmer velocity $U$ in the downward vertical direction as
 \begin{align}
 \label{eq:Squirmer_Speed_U}
     %U &= 2\frac{{\cal B}_1(\alpha)}{\lambda_B(\alpha)}B_1 - \frac{8}{3}\frac{{\cal B}_2(\alpha)}{\lambda_B(\alpha)}B_2 + \frac{g \triangle \rho R^2}{6\mu_f}\frac{1}{\lambda_B(\alpha)},\\
     U &= 2\frac{{\cal B}_1(\zeta)}{\lambda_B(\zeta)}B_1 - \frac{8}{3}\frac{{\cal B}_2(\zeta)}{\lambda_B(\zeta)}B_2 + \frac{F^\mathrm{grav}-F^\mathrm{rep}}{8\pi\mu_f R\lambda_B(\zeta)},\\
     \lambda_B(\zeta) &= \sinh \zeta \sum^{\infty}_{n=1}\frac{n(n+1)}{(2n-1)(2n+3)\triangle_n}\left[2\sinh (2n+1)\zeta + (2n+1)\sinh 2\zeta-\triangle_n\right],\\
     {\cal B}_1\left(\zeta\right) &= \sinh^3\zeta \sum^{\infty}_{n=1}\frac{n(n+1)}{\triangle_n}\left[1-\cosh(2n+1)\zeta+\sinh(2n+1)\zeta\right],\\
     {\cal B}_2\left(\zeta\right) &=
     \sinh^2\zeta \sum^{\infty}_{n=1}\frac{n(n+1)}{(2n-1)(2n+3)}\sinh\left(\left(n+\frac{1}{2}\right)\zeta\right)\frac{M_2(n,\zeta)}{\triangle_n},\\
     M_2\left(n,\zeta\right) &= \sinh\zeta(n-1)(n+2)\left[(2n+3)e^{-\left(n-\frac{1}{2}\right)\zeta}-(2n-1)e^{-\left(n+\frac{3}{2}\right)\zeta}\right] - \\
     & \frac{5}{4}\left[(n-1)(2n+3)e^{-\left(n-\frac{3}{2}\right)\zeta}+(2n+1)e^{-\left(n+\frac{1}{2}\right)\zeta}-(n+2)(2n-1)e^{-\left(n+\frac{5}{2}\right)\zeta}\right], \nonumber\\
     \triangle_n &= 4\sinh^2\left(n+\frac{1}{2}\right)\zeta - (2n+1)^2\sinh^2\zeta.
 \end{align}
 %with $\zeta = \acosh\left(h/R\right)$. 
 The drag coefficient for vertical motion in the presence of the no-slip wall is $c_D(\zeta)=8\pi\mu_f\lambda(\zeta)$ and the individual contributions to the net velocity from $B_1$, $B_2$, gravity, and repulsion are identified as
\begin{align}
     U_1 &= 2\frac{{\cal B}_1(\zeta)}{\lambda_B(\zeta)}B_1,\label{eq:U_1}\\
     U_2 &= - \frac{8}{3}\frac{{\cal B}_2(\zeta)}{\lambda_B(\zeta)}B_2,\label{eq:U_2}\\
     U^\mathrm{grav} &= \frac{F^\mathrm{grav}}{8\pi\mu_f R\lambda_B(\zeta)} =  V_g\cdot \frac{3}{4\lambda_B(\zeta)},\label{eq:U_grav}\\ 
     U^\mathrm{rep} &= -\frac{F^\mathrm{rep}}{8\pi\mu_f R\lambda_B(\zeta)},\label{eq:U_rep}
\end{align}
respectively, where $V_g=2 g\triangle \rho R^2/9\mu_f$ is the terminal velocity of the passive sphere in free space. 
We remark that in the situation where the squirmer is pointing vertically upwards ($\theta = +\pi/2$), the velocities (still expressed in the downward direction) are unchanged from the corresponding formulas in \eqref{eq:U_1}--\eqref{eq:U_rep} apart from a change in sign in \eqref{eq:U_1}.

\subsection{Far-field expansion of $U$ for a squirmer near a wall \label{subsec:far-field}}
The expression for the squirmer speed in Eq.~(\ref{eq:Squirmer_Speed_U}) allows for a far-field expansion of $U$ in the limit of $\zeta\rightarrow \infty$. In the absence of squirming activity ($B_1 = B_2 = 0$), the speed of a squirmer under gravity is
\begin{align}
\label{eq:U_g}
    U^\mathrm{grav} &=V_g \left(1-\frac{9}{8}\frac{R}{h} + \frac{1}{2} \left(\frac{R}{h}\right)^3-\frac{135}{256}\left(\frac{R}{h}\right)^4 - \frac{1}{8}\left(\frac{R}{h}\right)^5 + \frac{401}{512}\left(\frac{R}{h}\right)^6-\frac{675}{1024}\left(\frac{R}{h}\right)^7\cdots\right).
\end{align}
Eq.~(\ref{eq:U_g}) is identical, up to ${\cal O}\left((R/h)^3\right)$, to the often-used expression derived from the method of images with one image. The activity on the squirmer surface ($B_1$ and $B_2$) contributes to the far-field squirmer speed as
 \begin{align}
 \label{eq:U_g_B1_B2}
     U &=U^\mathrm{grav} + B_1\left(\frac{2}{3} - \frac{1}{3}\left(\frac{R}{h}\right)^3 + \frac{1}{6}\left(\frac{R}{h}\right)^5 -\frac{45}{128}\left(\frac{R}{h}\right)^6+\cdots\right) + \\
     & B_2 \left(\frac{3}{8}\left(\frac{R}{h}\right)^2 - \frac{1}{2}\left(\frac{R}{h}\right)^4 + \frac{15}{32}\left(\frac{R}{h}\right)^5 + \frac{5}{24}\left(\frac{R}{h}\right)^6 -  \frac{441}{512}\left(\frac{R}{h}\right)^7 + \cdots\right). \nonumber
 \end{align}

\begin{figure}
    \centering
    \includegraphics[width=1\textwidth]{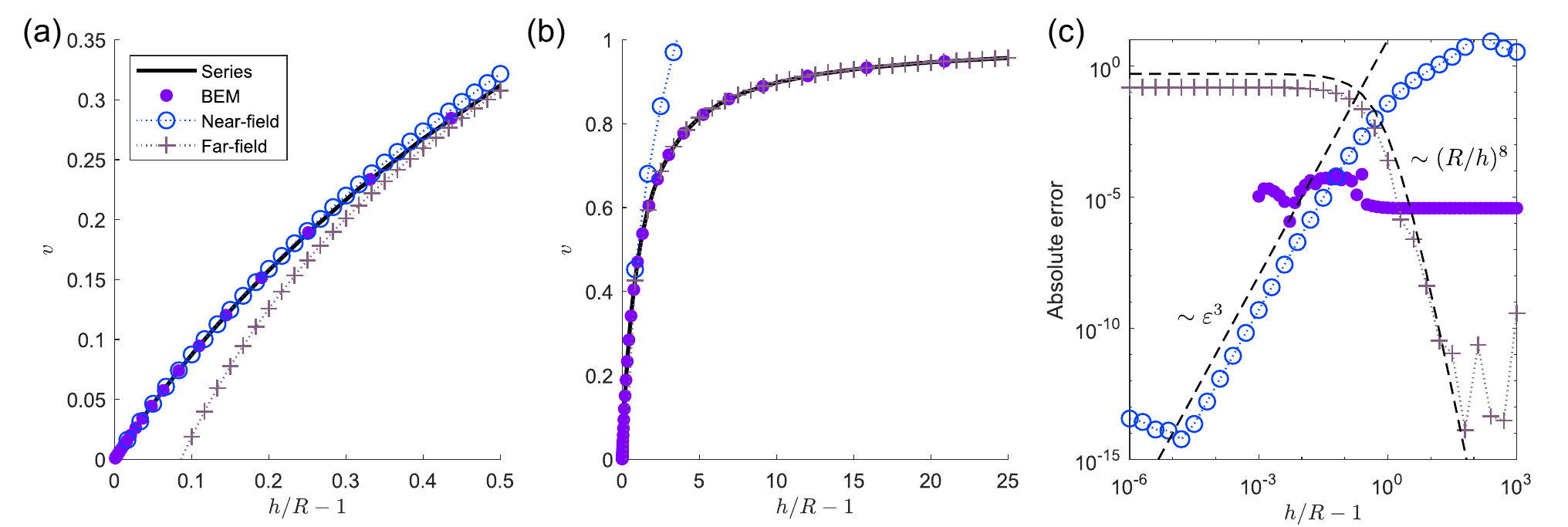}
    \caption{Comparison of the series expression \eqref{eq:Squirmer_Speed_U} with series truncated to 5000 terms, boundary integral simulation solutions, the near-field formula~\eqref{eq:Cooley_NearField}, and the far-field formula~\eqref{eq:U_g} for the vertical speed $v=U^\mathrm{grav}/V_g$ of a passive sphere ($B_1 = B_2 = 0$) sedimenting under gravity  near a no-slip wall without repulsion. Parts (a) and (b) display the same quantities focusing over different ranges of separation from the wall. (c) Errors with respect to the series solution.}
\label{fig:perp_vel_comparison_grav}
\end{figure}

\begin{figure}
    \centering
    \includegraphics[width=1\textwidth]{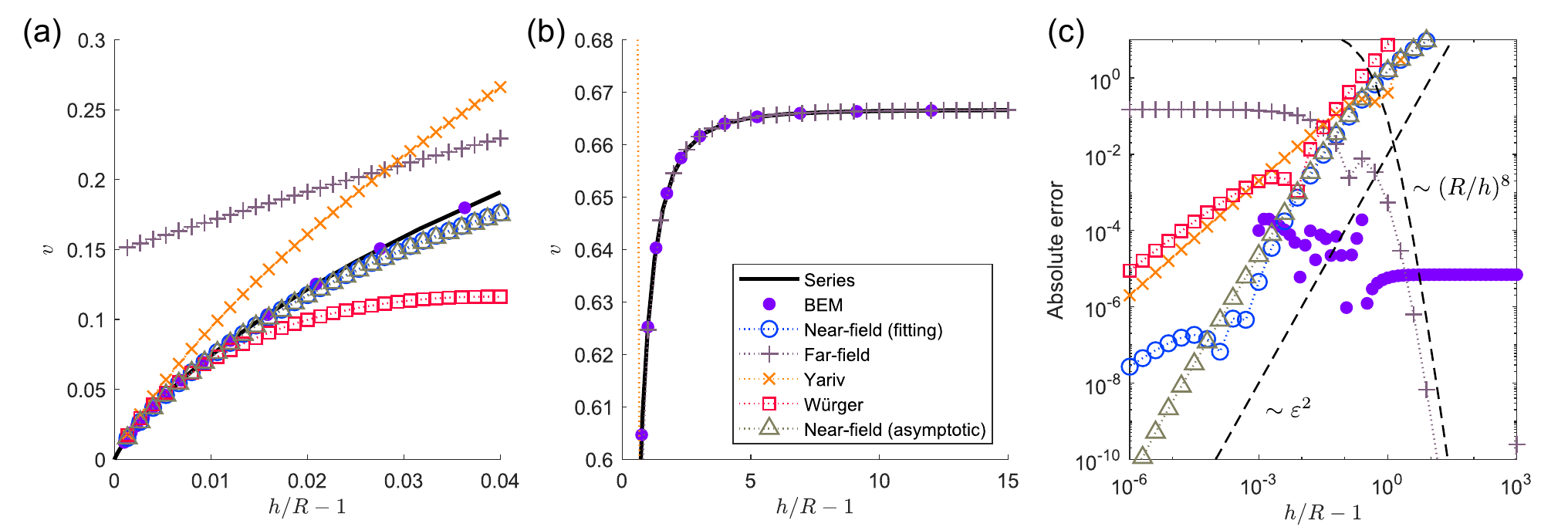}
    \caption{Comparison of the series expression \eqref{eq:Squirmer_Speed_U} with series truncated to 5000 terms, boundary integral simulation solutions, the near-field formula~\eqref{eq:our_NearField}, and the far-field formula \eqref{eq:U_g_B1_B2} for the vertical speed of a neutral squirmer ($B_1 = 1, B_2 = 0, V_g=0$) perpendicular to and near a no-slip wall without repulsion. Parts (a) and (b) display the same quantities focusing over different ranges of separation from the wall. (c) Errors with respect to the series solution.}
    \label{fig:perp_vel_comparison_B1}
\end{figure}

\begin{figure}
    \centering
    \includegraphics[width=1\textwidth]{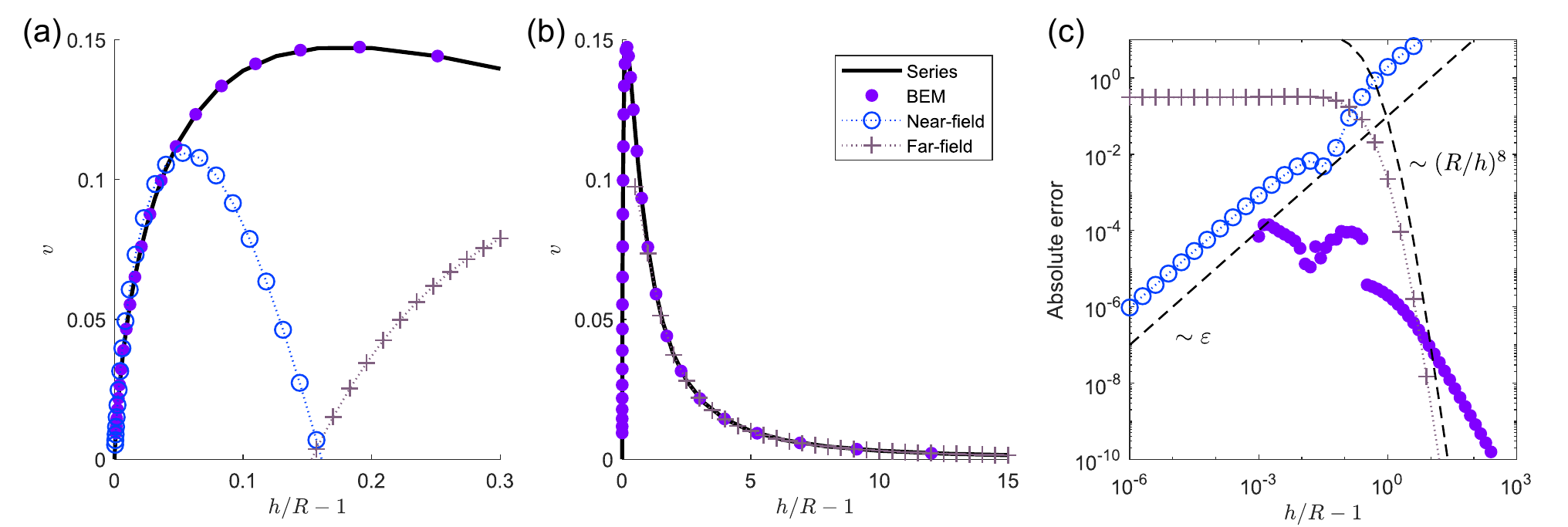}
    \caption{Comparison of the series expression \eqref{eq:Squirmer_Speed_U} with series truncated to 5000 terms, boundary integral simulation solutions, the near-field formula~\eqref{eq:our_NearField}, and the far-field formula \eqref{eq:U_g_B1_B2} for the vertical speed of a contractile squirmer ($B_1 = 0, B_2 = 1, V_g=0$) perpendicular to and near a no-slip wall without repulsion. Parts (a) and (b) display the same quantities focusing over different ranges of separation from the wall. (c) Errors with respect to the series solution.}
    \label{fig:perp_vel_comparison_B2}
\end{figure}

\subsection{Near-field $U$ for a squirmer near a wall \label{subsec:near-field}} 
\cite{cox1967slow} and \cite{cooley1969slow} provided an expression for the near-field velocity of a sedimenting rigid sphere ($B_1 = B_2 = 0$) at a distance $h$ above a planar wall, with $\varepsilon = h/R-1 \ll 1$ and $\zeta = \acosh\left(1+\varepsilon\right)$,
\begin{align}
    \label{eq:Cooley_NearField}
    U^\mathrm{grav} &= \frac{\tfrac{4}{3}\pi R^3\Delta \rho g}{6 \pi \mu_f R}\frac{1}{\varepsilon^{-1}-\left(0.2\ln\varepsilon - 0.971280\right)} \\
    &=V_g \frac{1}{\varepsilon^{-1}-\left(0.2\ln\varepsilon - 0.971280\right)}.
\end{align}
This result shows that the infinite series $\lambda_B$ in the denominators on the right-hand-side of Eq.(\ref{eq:Squirmer_Speed_U}) has the following asymptotic behavior as $\varepsilon \rightarrow 0$:   
\begin{align}
    \label{eq:lambdaB_NearField} \lambda_B\left(\zeta(\varepsilon)\right) &\rightarrow \overline{\lambda_B} \equiv \frac{3}{4}\left(\varepsilon^{-1}-0.2\ln{\varepsilon}+0.971280\right).
\end{align}
This is confirmed in Fig.~(\ref{fig:perp_vel_comparison_grav}), showing both the truncated series  solution and the boundary integral simulation results  are in good agreement with the near-field asymptotic expression in Eq.~(\ref{eq:Cooley_NearField}).

For a squirmer with a prescribed slip velocity $B_1\ne 0$ and $B_2=0$ moving towards the wall ($\theta=-\pi/2$), \cite{yariv2016thermophoresis} provided an asymptotic expression for the near-field velocity (adapted to our variable definitions)
\begin{align}
\label{eq:Yariv_NearField}
    U_1&\approx -2B_1 \varepsilon\left(\ln \varepsilon - 0.1087\right),
\end{align}
which corresponds to
\begin{align}
\label{eq:Yariv_NearField_B1}
    {\cal B}_1(\zeta(\varepsilon)) &\approx -\frac{3}{4}(\ln\varepsilon - 0.1087).
\end{align}
\citet{wurger2016hydrodynamic} considered a simplifying assumption and derived an asymptotic result where the constant $-0.1087$ in \eqref{eq:Yariv_NearField} is replaced by $+2.25$. Based on the results in Eqs.~(\ref{eq:lambdaB_NearField})-(\ref{eq:Yariv_NearField_B1}), we construct an ansatz for the near-field behavior of the squirmer speed in Eq.~(\ref{eq:Squirmer_Speed_U}), assuming identical forms for the approximations to ${\cal B}_1$ and ${\cal B}_2$, with $\varepsilon\ll 1$: 
\begin{align}
\label{eq:our_NearField}
    U &\approx \frac{1}{\overline{\lambda_B}}\left[ 2B_1a_1\left(\ln\varepsilon + b_1\right)-\frac{8}{3}B_2 a_2\left(\ln\varepsilon+b_2\right) + \frac{3}{4}V_g\right],
\end{align}
We then compute the coefficients $a_1, b_1, a_2,$ and $b_2$ by least-squares fitting of the series solution truncated at $5000$ terms over the range ${10^{-5}\leq \varepsilon \leq 10^{-3}}$ to Eq.~(\ref{eq:our_NearField}), and obtain: 
\begin{equation}
    \label{eq:least_squares_fit_values}
    a_1 = -0.7476,\quad b_1 = 0.8633, \quad a_2 = 1.006, \quad b_2 = 1.825. 
\end{equation}
As an independent check, 
we apply the methods in \cite{cox1967slow} to compute the near-field asymptotic expansion for a squirmer with $B_1=1$ and  $B_2=0$, and find $a_1=-0.75$ and $b_1=0.8913$, in good agreement with the values from the least-squares fitting. 

We next compare the asymptotic near-field results from both \citep{yariv2016thermophoresis} and \citep{wurger2016hydrodynamic} with our boundary integral simulation results and our near-field expression in Eq.~(\ref{eq:our_NearField}) in Fig.~(\ref{fig:perp_vel_comparison_B1}) with $B_1=1$ and $B_2=0$. We find that our near-field expression in Eq.~(\ref{eq:our_NearField})-(\ref{eq:least_squares_fit_values}) is close to the truncated series solution for $10^{-6}\leq \varepsilon\leq 0.05$ and close to the boundary integral simulation for $10^{-3}\leq \varepsilon\leq 0.05$ (the boundary integral solution is not accurate for $\varepsilon<10^{-3}$), while the asymptotic expressions from both \cite{yariv2016thermophoresis} and \cite{wurger2016hydrodynamic} diverge significantly for $\varepsilon \ge 0.02$. 
For $B_1=0$ and $B_2=1$ we compare our near-field expression with the truncated series and boundary integral simulation results in Fig.~(\ref{fig:perp_vel_comparison_B2}). 

 \begin{figure}
    \centering

    \includegraphics[scale=0.75]{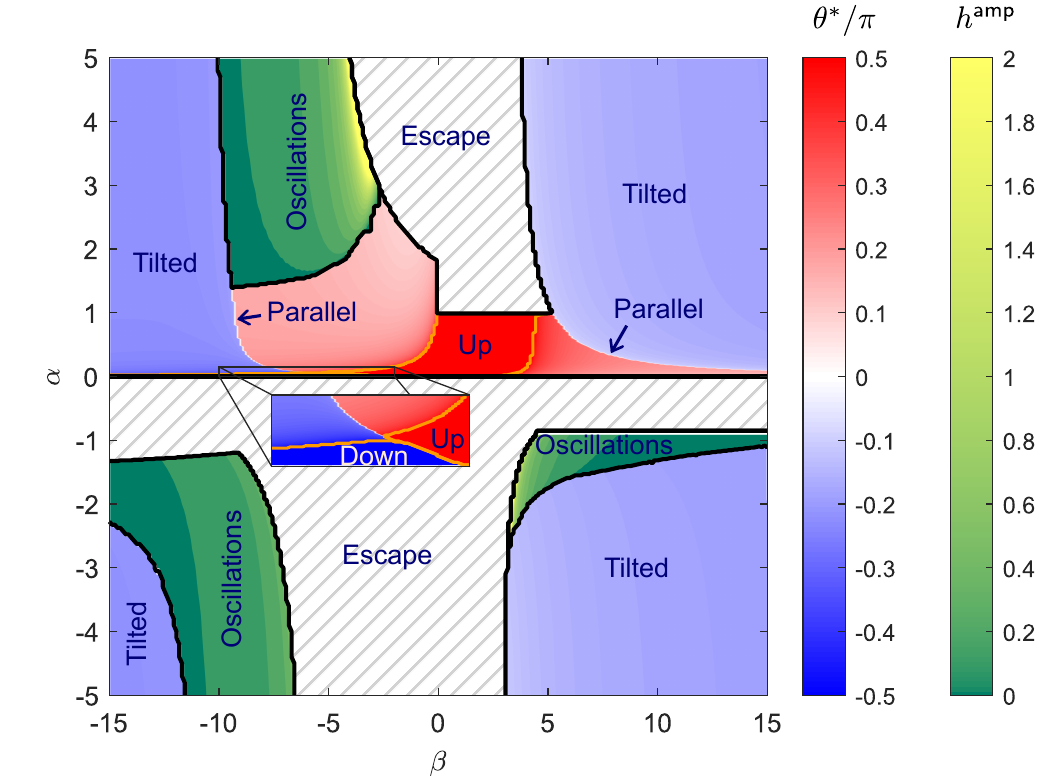}
    
    \caption{Long-time behaviors of a squirmer under gravity next to a flat wall, with the squirmer initially pointing nearly vertically downwards ($\theta(t=0)=-0.99\pi/2$) at a starting height $h/R=10$. The squirmer is not bound to the wall in the ``Escape" region. For squirmers bound to the wall under gravity, they settle to either a steady state at a fixed height with a steady tilt angle sliding along the wall (red--blue color bar), or they oscillate  
    in the ``Oscillations" region where the squirmer-wall distance oscillates with amplitudes in height indicated by the right (green--yellow) color bar. Negative values of $\alpha$ signify that gravity acts vertically away from the wall. The inset shows the detailed distribution of swimming dynamics for $0<\alpha<0.15$ and $-10<\beta<-2$.}
    \label{fig:phase_map_initial_down}
\end{figure}

\section{Swimming dynamics of a squirmer under gravity near a no-slip wall\label{sec:general_swimming_dynamics}} 
\subsection{Classification of squirmer dynamics in the $\alpha-\beta$ plane \label{subsec:mixed_modes}}
%%%%%%%%%%%%%%%%%%%%%%%%%%%%%%%%%%%%%%%%%%%%%%%%%%%%%%%%%%%%%%%%%%%%%%%
%%%%%%%%%%%%%%%%%%%%%%%%%%%%%%%%%%%%%%%%%%%%%%%%%%%%%%%%%%%%%%%%%%%%%%%
Under gravity, various types of swimming dynamics arise from 
the hydrodynamic interaction between a squirmer and a no-slip flat wall \citep{li2014hydrodynamic,lintuvuori2016hydrodynamic,ruhle2018gravity,thery2023hydrodynamic}. We first define two parameters to quantify such diverse swimming dynamics: (i) $\alpha=V/V_g$, the ratio of self-propulsion speed due to $B_1$ mode to the gravity-induced speed, and (ii) $\beta=B_2/B_1$, the ratio of the two squirming mode magnitudes.

For $\alpha>1$, the squirmer is prone to escape from the wall in the long-time limit, even though its initial height and orientation may lead to a transient contact with the wall. We consider a trajectory to have escaped the wall if $h/R > 100$ at any positive time. For a squirmer that is bound to the wall, at least three types of swimming dynamics have been reported:   (1) The squirmer is pinned close to the wall at a fixed height, pointing either toward or away from the wall, depending on the values of $(\alpha,\beta)$. (2) The squirmer slides along the wall at a fixed height with a tilted orientation. (3) The squirmer oscillates (bounces) in both height and orientation. While these near-wall swimming dynamics under gravity have been reported in the literature \citep{li2014hydrodynamic,ruhle2018gravity}, no detailed investigation on how these states may bifurcate in terms of $(\alpha,\beta)$ is available to our knowledge. 
\begin{figure}
    \centering
    
    \includegraphics[scale=0.75]{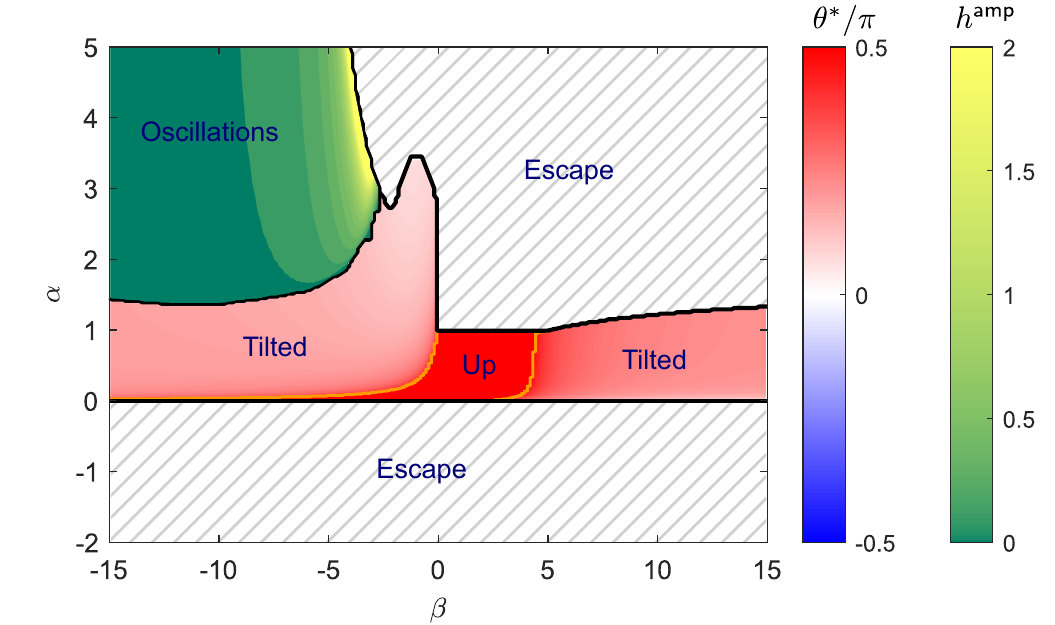}
    
    \caption{Long-time behaviors of a squirmer under gravity next to a flat wall, with the squirmer initially pointing horizontally (parallel to the wall) at a starting height $h/R=10$. The squirmer is not bound to the wall in the ``Escape" region. For squirmers bound to the wall under gravity, they  settle to either a steady state at a fixed height with a steady tilt angle (red--blue color bar), or they oscillate  
    in the ``Oscillations" region where the squirmer-wall distance oscillates with amplitudes in height indicated by the right (green--yellow) color bar. Negative values of $\alpha$ signify that gravity acts vertically away from the wall.}
    \label{fig:phase_map_parallel}
\end{figure}
\begin{figure}
    \centering
    
    \includegraphics[scale=0.75]{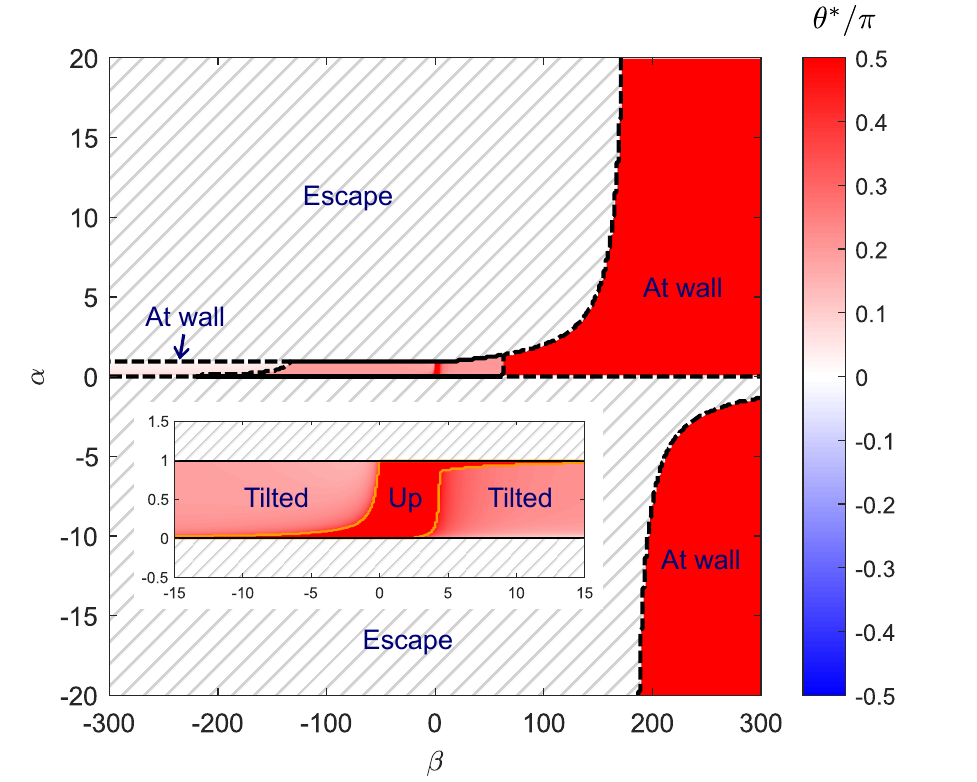}
    
    \caption{Long-time behaviors of a squirmer under gravity next to a flat wall, with the squirmer initially pointing nearly vertically upwards ($\theta(t=0)=0.99\pi/2$) at a starting height $h/R=10$. The squirmer is not bound to the wall in the ``Escape" region. For squirmers bound to the wall under gravity, they  settle to either a steady state at a fixed height with a steady tilt angle indicated by the color bar. In regions labelled as ``at wall,'' the squirmer approaches the minimum wall separation for which velocities were computed so trajectories could not be continued further in time. Negative values of $\alpha$ signify that gravity acts vertically away from the wall. An inset shows more detail for the range $-0.5<\alpha<1.5$, $-15<\beta<15$.}
    \label{fig:phase_map_up}
\end{figure}

\begin{figure}
    \centering

    \includegraphics[trim=0cm 0 0cm 0,clip,width=\textwidth]{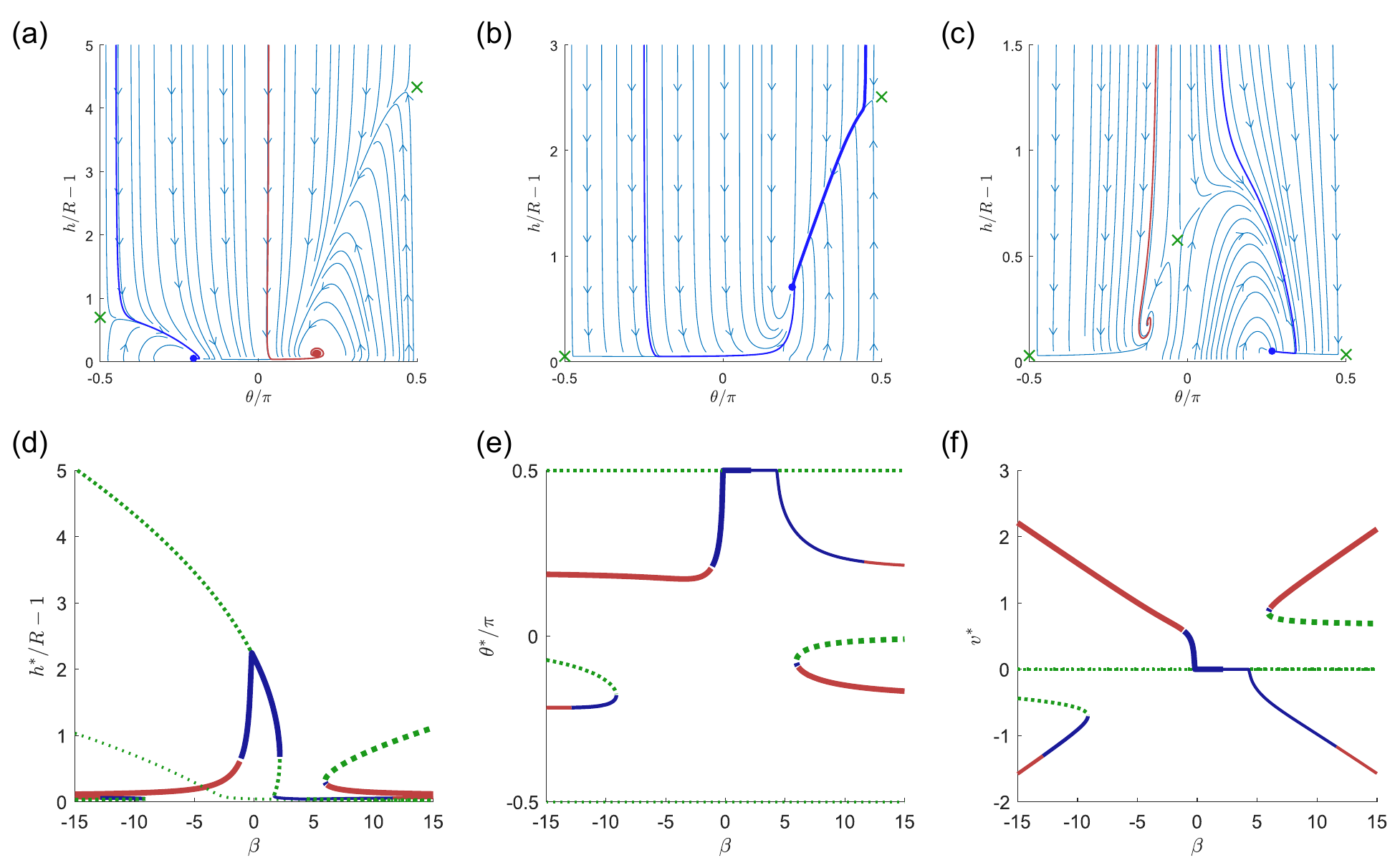}
    
    \caption{Bifurcation structures of swimming dynamics of a squirmer with $\alpha=2/3$. Top row: (a) $\beta=-10$, with two saddles (crosses), a stable node (blue filled circle), and a stable spiral, (b) $\beta=-1$, with two saddles and a stable node, and (c) $\beta=7.5$, with three saddles, a stable node, and a stable spiral. Bottom row: Dependence of (d) stable height, (e) orientation, and (f) wall-parallel speed on $\beta$. Red curves indicate stable spirals and blue curves indicate stable nodes. Green dashed curves indicates saddle nodes. Different curve thicknesses are used to visually distinguish branches of stationary points.}
    \label{fig:bifurcation_alpha_23rds}
\end{figure}
\begin{figure}
    \centering

    \includegraphics[width=\textwidth]{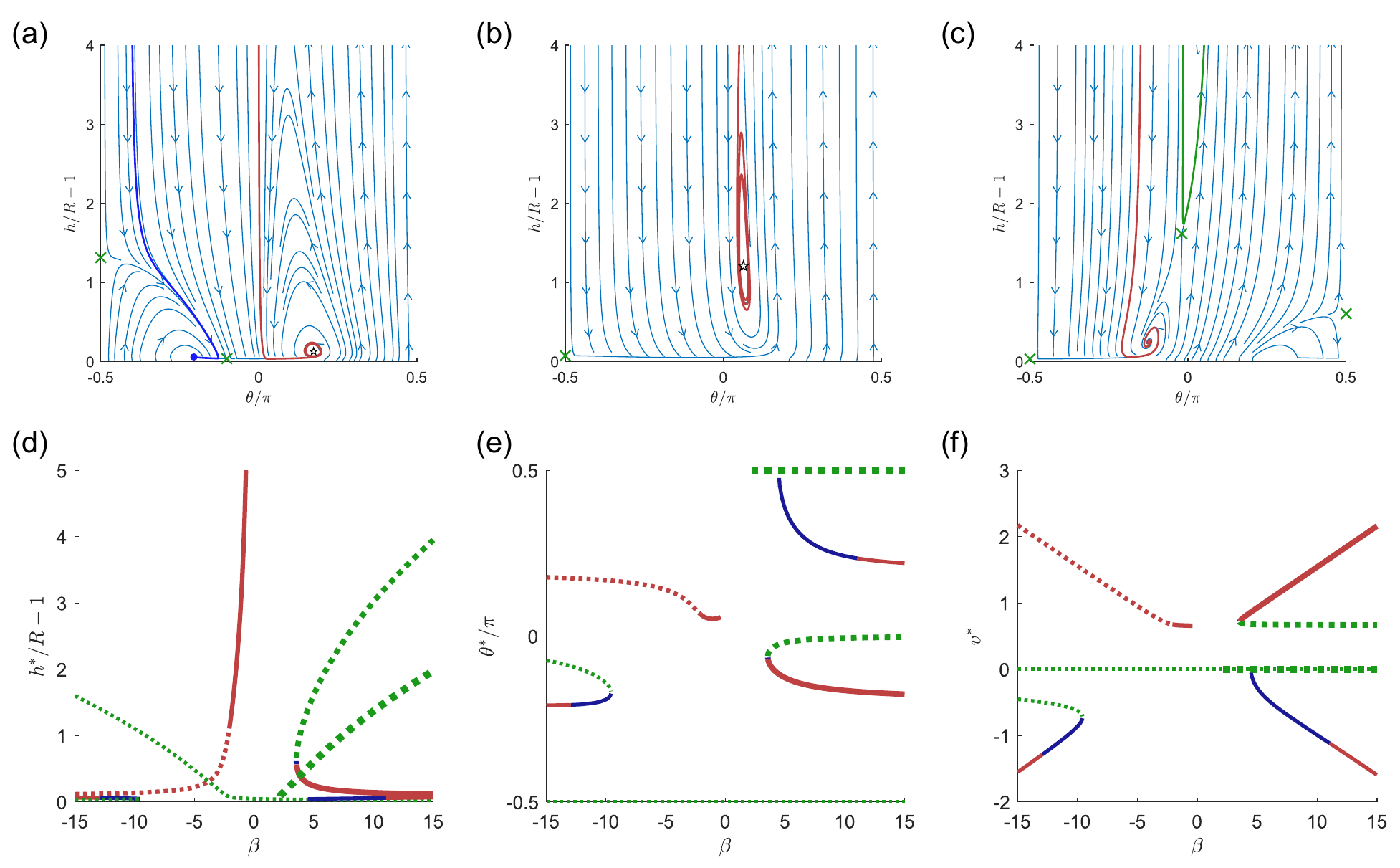}
    
    \caption{Bifurcation structures of swimming dynamics of a squirmer with $\alpha = 5$. %\rho=2/15$. 
    Top row: (a) $\beta=-12$, with two saddles (crosses), a stable node (filled circle), and an unstable spiral (star) enclosed by a limit cycle, (b) $\beta=-2$, with a saddle and a stable spiral (for clarity, only a portion of a trajectory attracted to the spiral is shown), and (c) $\beta=5$, with three saddles and a stable spiral. Bottom row: Dependence of (d) stable height, (e) orientation, and  (f) wall-parallel speed on $\beta$. Red curves indicate stable spirals and blue curves indicate stable nodes. Green dashed curves indicates saddle nodes. Different curve thicknesses are used to visually distinguish branches of stationary points.}
    \label{fig:bifurcation_alpha_five}
\end{figure}
\begin{figure}
    \centering
    
    \includegraphics[width=0.9\textwidth]{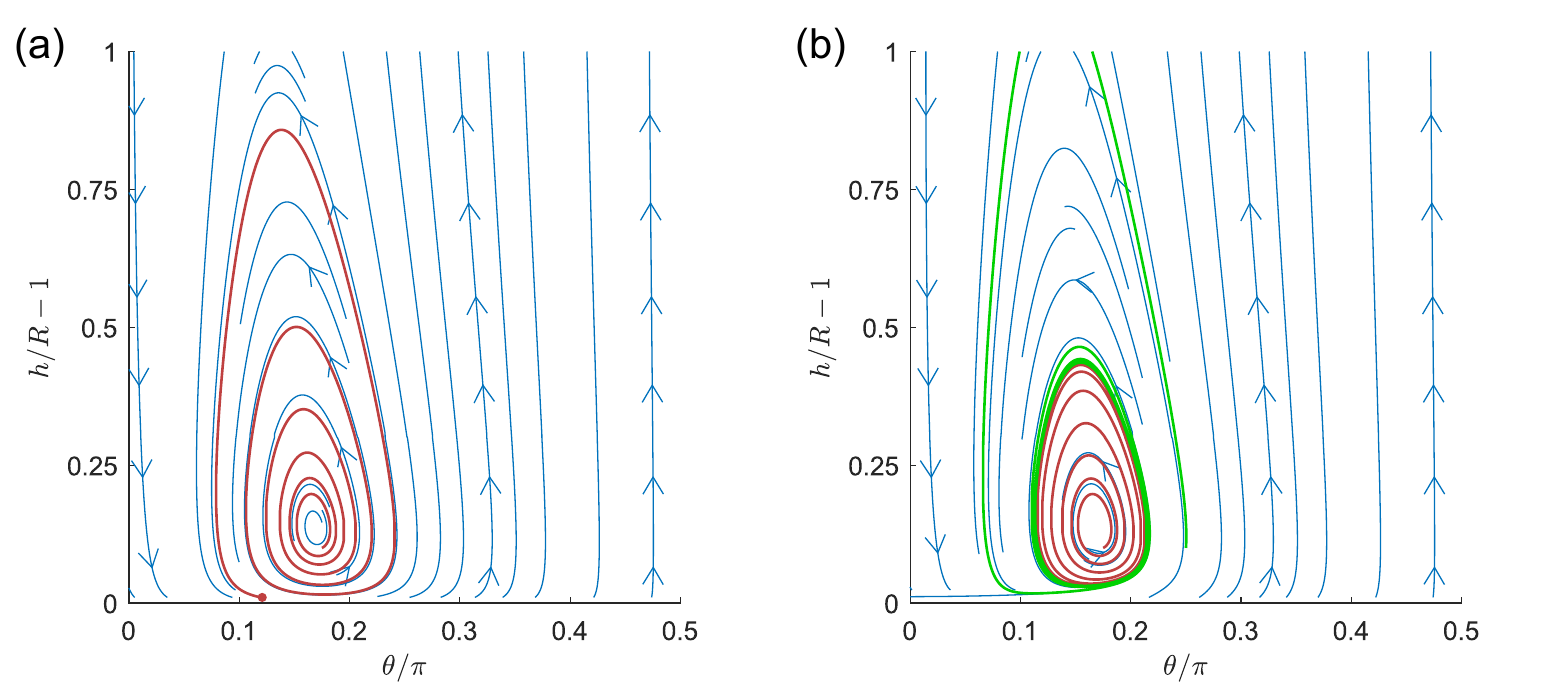}    
    \caption{Effect of repulsion on the swimming dynamics of a squirmer with $\beta=-10$ $(B_1=1, B_2 = -10)$, and $\alpha = 5$. %\rho=2/15$
    Phase plane dynamics near an unstable spiral point (a) with no wall repulsion, and (b) with short-range wall repulsion given in Eq.~\ref{eq:repulsion}. Red curves show trajectories starting close to the spiral point. The red dot in (a) (with no repulsion) indicates where the trajectory terminates due to close proximity with the wall. With repulsion in (b), the trajectory approaches a limit cycle, in which the distance from the wall and the orientation of the squirmer oscillate periodically. A trajectory starting outside this limit cycle is also shown (green).}
    \label{fig:oscillation_alpha_five}
\end{figure}

%%%%%%%%%%%%%%%%%%%%%%%%%%%%%%%%%%%%%%%%%

From both numerical simulations and far-field analysis, we find that the squirmer is less likely to be bound to the wall if its initial orientation points away from the wall. Thus we first focus on the various dynamics of a squirmer initially pointing towards the flat wall. We use the efficient and accurate boundary integral codes to evaluate the velocities of the squirmer on a grid in $(\theta,h)$ configuration space. We then compute streamlines to map out the various squirming dynamics in the $(\alpha,\beta)$-plane for a squirmer initially pointing nearly vertically down (specifically, the initial orientation angle is $\theta_0=-0.99\pi/2$) at a height of $h_0=10R$. The various swimming dynamics is summarized in Fig.~(\ref{fig:phase_map_initial_down}). For pinned and sliding dynamics we color code their regions using the orientation angle, with $\theta^*=\pi/2$ pointing up (red) and $\theta^*=-\pi/2$ pointing down (blue). The region for oscillatory dynamics is color coded by the oscillation amplitude ($h^{\text{amp}}$) of squirmer height (green). We note that while oscillations can be obtained over a large region of parameter space, the amplitudes are small except near the boundary with escaping behavior. Fig.~(\ref{fig:phase_map_parallel}) shows the distribution of swimming dynamics for a squirmer initially parallel to the wall.  For a squirmer initially pointing away from the wall (vertically up), Fig.~(\ref{fig:phase_map_up}) shows that the squirmer can stay bound to the wall for a wider range of $\alpha$ for $\beta\gg 1$, when the upward pointing squirmer moves toward the wall and stays bound to the wall for $|\alpha| \gg 1$ due to the strong puller mode $B_2 \gg B_1$.

The squirmer dynamics summarized in Figs.~(\ref{fig:phase_map_initial_down}-\ref{fig:phase_map_up}) offer some general observations: (1) A squirmer stays bound to the bottom wall under gravity as long as $\alpha$ is in the range $\alpha \in [0,1]$ for all values of $\beta$. (2) The oscillatory dynamics of a squirmer under gravity is generally associated with negative $\beta$ (extensile squirmers). When gravity is directed away from the wall (for example, if the wall is at the top of a chamber), however, it is possible to observe oscillatory dynamics for both contractile and extensile squirmers. (3) For a squirmer initially pointed towards the wall  the squirmer can be pinned facing down, albeit in a very small region (insert of Fig.~(\ref{fig:phase_map_initial_down})). (4) The squirmer does not bounce (oscillate) around the wall if it initially points away from the wall. (5) For sufficiently large $\beta$ we expect that a squirmer can be bound to the wall, independent of squirmer's initial configuration and the value of $\alpha>0$.

We next examine the detailed bifurcation structures of squirmer swimming dynamics as a function of $\beta$ for various fixed values of $\alpha$. An example is shown in Fig.~(\ref{fig:bifurcation_alpha_23rds}), where $\alpha = 2/3$ and we vary $\beta=B_2/B_1$ with $B_1=1$ and $C^\mathrm{rep}=10^4$. We quantify the swimming dynamics via the equilibrium states (the fixed points) in the flow map of the squirmer in the plane of height versus angle. In the top row of Fig.~(\ref{fig:bifurcation_alpha_23rds}), the red curves are for stable spirals and blue curves are for stable nodes in the flow map. A stable spiral and a stable node are found for $\beta=-10$ in (a) and $\beta=7.5$ in (c). In (b), only one stable node is found for $\beta=-1$. The bottom row summarizes the bifurcation structures for $\beta=B_2/B_1 \in [-15,15]$, with the dotted curves denoting the unstable branches.

At a higher squirmer propulsion speed (higher value of $\alpha$), we expect the region for stable nodes to shrink. An example with $\alpha=5$ is shown in Fig.~(\ref{fig:bifurcation_alpha_five}). In the top row, the phase plane diagrams are shown for $\beta=-12$ in (a), $-2$ in (b), and $5$ in (c). For $\beta=-12$, squirmers that approach the wall are attracted either to a stable node with a negative tilt or to a limit cycle with positive tilt; both of these attractors are very close to the wall. For $\beta=-2$, most initial conditions lead to escape from the wall but some are attracted to a stable spiral. For $\beta=5$ (c), the squirmer either spirals into a fixed point close to the wall or escapes, depending on the initial angle. There is an additional stable node close to the wall that cannot be reached by trajectories that begin far from the wall. The bifurcations with respect to $\beta$ are summarized in the bottom row of Fig.~(\ref{fig:bifurcation_alpha_five}). We find that the branch of stable nodes that was present for $\alpha=2/3$ (blue curves in Fig.~(\ref{fig:bifurcation_alpha_23rds})) at intermediate values of $\beta$ disappear at $\alpha=5$. In fact, we find no stable equilibrium points for $\beta\in(-9.5,-2.07)\cup(0,3.5)$. For $\beta<-2.07$, there is an unstable spiral around which we expect a limit cycle (as shown in Fig.~(\ref{fig:bifurcation_alpha_five}a)) and for $\beta\in(0,3.5)$, the generic behavior is to escape from the wall.  

Fig.~(\ref{fig:oscillation_alpha_five}) shows the effect of wall repulsion on the spiraling squirmer dynamics, which transitions from (a) an unstable spiral that eventually becomes too close to the wall for numerical solutions to continue into (b) a limit cycle, giving rise to oscillatory dynamics of a squirmer in both height and orientation due to a wall repulsion that is strong enough to maintain a minimum separation between the wall and the squirmer.

\begin{figure}
    \centering
    \includegraphics[width=\textwidth]{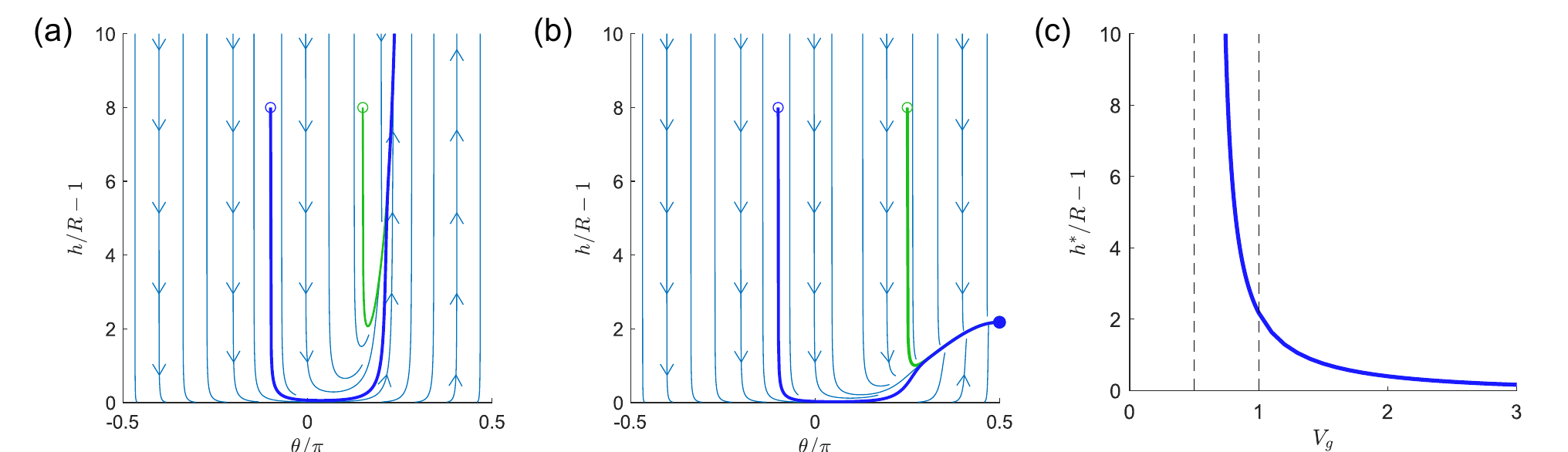}
    \caption{Threshold in $V_g$ for a steady equilibrium height and angle for a squirmer with $B_1=1, B_2=0$ (neutral swimmer). (a) Escaping dynamics for a squirmer with $V_g=0.5$. (b) A squirmer with $V_g=1$ reaches a steady equilibrium height and points upward. (c) Dependence of stable height in perpendicular-up orientation on the free space sedimentation speed $V_g$. Dashed vertical lines indicate the values of $V_g$ used in (a) and (b) respectively.}
    \label{fig:B11_B20}
\end{figure}
\begin{figure}
    \centering
    
    \includegraphics[width=\textwidth]{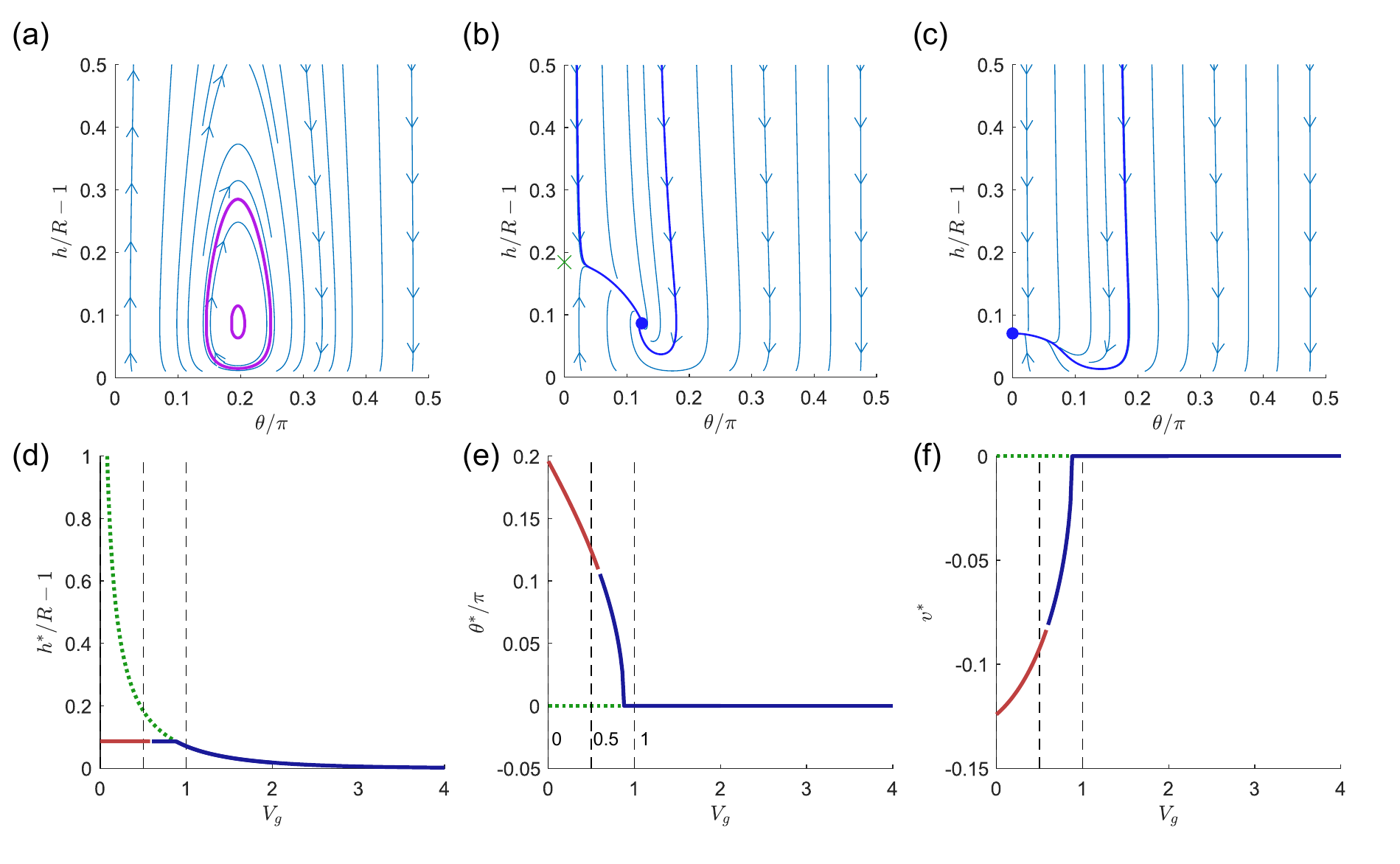}
    
    \caption{Bifurcation structures of swimming dynamics of a squirmer with $B_1=0, B_2=1$ (contractile shaker). Top row (a) $V_g=0$, (b) $V_g=0.5$, and (c) $V_g=1$. Bottom row: Dependence of (a) stable height, (b) orientation, and (c) wall-parallel speed on the free space sedimentation speed $V_g$. Red curves indicate stable spirals and blue curves indicate stable nodes. Green dashed curves indicates saddle nodes.}
    \label{fig:B10_B21}
\end{figure}

\subsection{Effects of gravity (varying $V_g$) for a pure squirmer \label{subsec:pure_modes}}
\begin{figure}
    \centering

    \includegraphics[width=\textwidth]{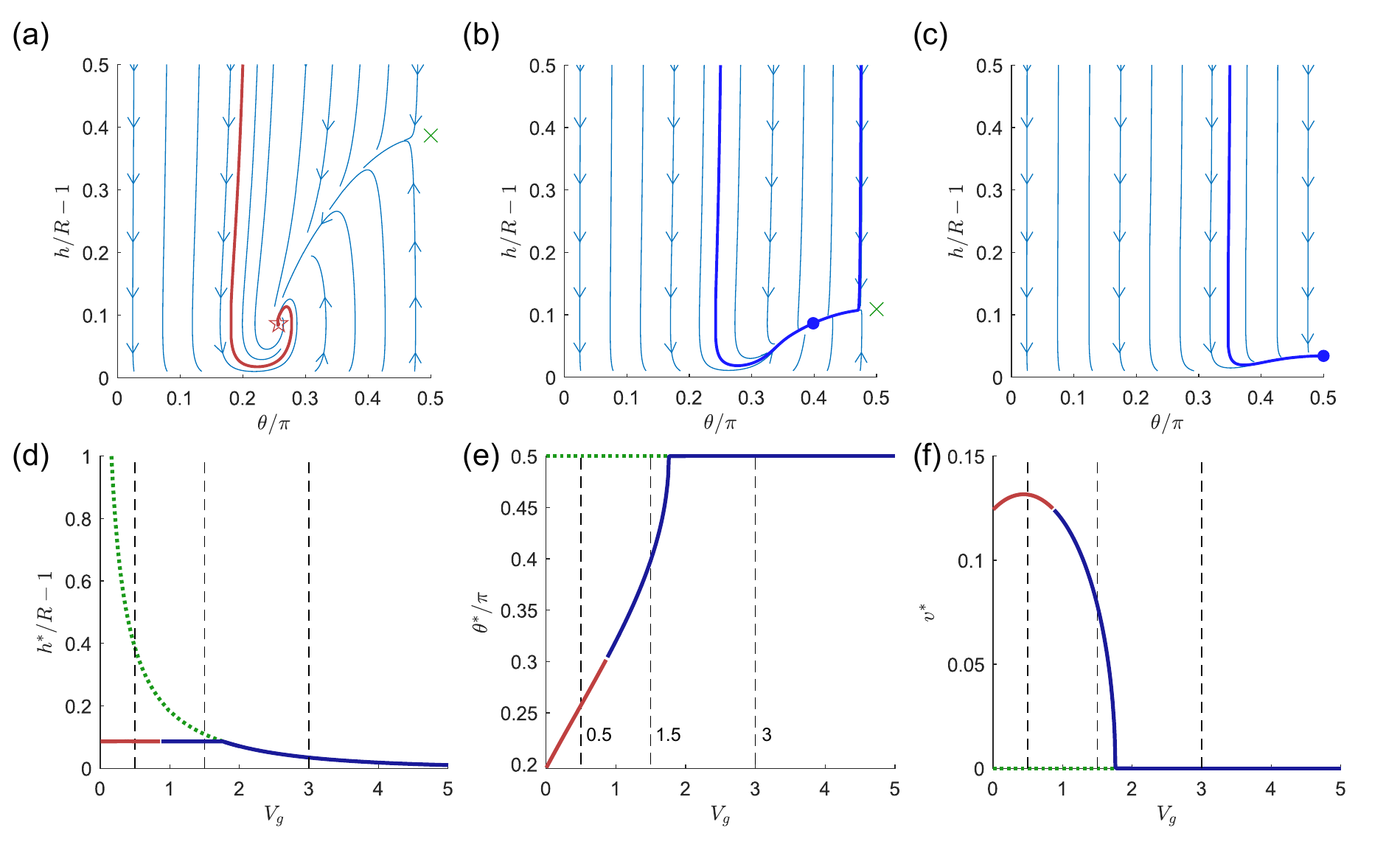}
    
    \caption{Bifurcation structures of swimming dynamics for a squirmer with $B_1=0, B_2=-1$ (extensile shaker). Top row:  (a) $V_g=0.5$, (b) $V_g=1.5$, and (c) $V_g=3$. Bottom row: Dependence of (d) stable height, (e) orientation, and (f) wall-parallel speed on the free space sedimentation speed $V_g$. Red curves indicate stable spirals and blue curves indicate stable nodes. Green dashed curves indicates saddle nodes.}
    \label{fig:B10_B2m1}
\end{figure}

Here we investigate the swimming dynamics of a pure squirmer (either a ``shaker" with $B_1=0$, $B_2 \neq 0$ or a ``neutral swimmer" with $B_1\neq 0$, $B_2=0$) under gravity, focusing on three combinations of $(B_1, B_2)$: $(1,0)$ for a neutral swimmer in Fig.~(\ref{fig:B11_B20}), $(0,1)$ for a contractile shaker (puller) in Fig.~(\ref{fig:B10_B21}) and $(0,-1)$ for an extensile shaker (pusher) in Fig.~(\ref{fig:B10_B2m1}), and examine the swimming dynamics as a function of $V_g$.

For a neutral swimmer with $B_1=1$ and $B_2=0$, we find that it can 
be bound to the wall and stay at an equilibrium height, pointing toward the wall with zero velocity ($v^*=0$) as long as the gravity is sufficiently large so that $V_g >2/3$, see Fig.~(\ref{fig:B11_B20}).

Slightly more complicated swimming dynamics is found for a contractile puller shaker ($B_1=0$ and $B_2=1$): We find a branch of saddle nodes where the squirmer points to the wall with zero velocity at a finite height. Co-existent with this saddle node is a stable spiral with a tilt angle and a sliding velocity for intermediate gravity. For sufficiently large $V_g$, the purely extensile squirmer is bound to the wall at a fixed height, pointing to the wall with zero velocity. Furthermore we find that for a contractile puller shaker under a sufficiently large $V_g$, the squirmer can reach an equilibrium height with its director parallel to the wall and a zero sliding velocity, see Fig.~(\ref{fig:B10_B21})(e) and (f), where the angle $\theta^*=0$ and $v^*=0$ for $V_g \ge 1$. 

The bifurcation structure of the swimming dynamics of an extensile pusher shaker is summarized in Fig.~(\ref{fig:B10_B2m1}), where we find that the steady state ($v^*=0$ for large $V_g$) is a squirmer at a steady equilibrium height while pointing upright, see Fig.~(\ref{fig:B10_B2m1})(e) and (f), where the angle $\theta^*=\pi/2$ and $v^*=0$ for $V_g \ge \sim 1.6$.

\section{Discussion and Conclusion \label{sec:conclusion}}
In this work we provide an exact solution for a spherical squirmer sedimenting to a flat solid wall. We provide both far-field and near-field approximations to the swimming velocity of a squirmer under gravity, and show that our near-field approximations, different from both \citep{yariv2016thermophoresis} and \citep{wurger2016hydrodynamic}, are valid over a wider range of squirmer distances to the wall. We next use boundary integral simulations to map out its various swimming dynamics in the $\alpha-\beta$ plane, and find that the squirmer may escape from the wall, slide along the wall at a fixed height and orientation, stay at a fixed height pointing to or away from the wall, or oscillate in both height and orientation. 

We further examine the bifurcations in the steady state configurations of the squirmer interacting with a solid wall as the parameters $\alpha$ and $\beta$ are varied, identifying branches of stable and unstable spirals, stable nodes, and saddle nodes in $\theta$--$h$ phase space. In particular, we find that there are parameter regions where different swimming dynamics coexist (overlaying solid branches for stable spiral and stable node in Figs.~(\ref{fig:bifurcation_alpha_23rds}-\ref{fig:bifurcation_alpha_five})). Such identification allows us to characterize when the squirmer will escape from the wall, remain bound to the wall at a fixed location, slide along the wall at a fixed height, or bounce along the wall, making it possible to design robotic microswimmers that can adjust their gaits (by changing the values of $B_1$ and $B_2$) to navigate along the solid wall in the presence of obstacles. 

{\color{black}  Particle--wall interactions in complex biological fluids often involve effects of non-Newtonian rheology. For example, biological fluids such as blood and mucus are typically shear-thinning fluids that have profound effects on locomotion of microswimmers. Novel particle--wall interactions in a shear-thinning non-Newtonian fluid have been studied in the context of the rolling of a rotating sphere near a solid wall \citep{chen2021wall}. For a sphere rolling on the wall, the non-Newtonian shear-thinning rheology can give rise to wall-induced translation opposite to the direction of friction against the wall \citep{chen2021wall}. \cite{li2017near} studied the effects of non-Newtonian rheology on the interactions between an undulatory swimmer next to a wall. How would non-Newtonian rheology alter the swimming dynamics of a squirming sphere next to a rigid wall? It would be interesting to investigate the non-Newtonian effects on the bifurcation structures of the swimming dynamics of a squirmer next to a solid wall.

In a porous medium, the interactions between the active swimmers and the complex boundaries are essential to understanding the diffusive transport of active suspensions such as bacteria that transition between states as they run and tumble \citep{datta2024random}. Our results show that, under gravity, the squirmer has multiple trajectories between being bound to the wall or escaping from the wall by varying $\alpha$ (the relative propulsion velocity to the sedimenting velocity) and $\beta$ (the relative strength of the swimming mode to the contractile/extensile mode). For a squirmer sedimenting in a porous medium, each time it encounters an obstacle it can transition between states that we reported here, similar to the transitions of run-and-tumbling bacteria in porous media \citep{mattingly2023bacterial}. It would be interesting to quantify how the effective diffusivity of squirmer sedimenting in a porous medium depends on $\alpha$ and $\beta$.}

\section*{Acknowledgement}
The authors acknowledge useful discussions with Bryan Quaife and On Shun Pak. YNY acknowledges support by the National Science Foundation (NSF) under award DMS- 1951600, and of the Flatiron Institute, part of Simons Foundation.
HS acknowledges the support of the Natural Sciences and Engineering Research Council of Canada (NSERC), [funding reference number RGPIN-2018-04418].
Cette recherche a \'{e}t\'{e} financ\'{e}e par le Conseil de recherches en sciences naturelles et en g\'{e}nie du Canada (CRSNG), [num\'{e}ro de r\'{e}f\'{e}rence RGPIN-2018-04418].

\noindent
{\bf Declaration of Interests: The authors report no conflict of interest.}

\bibliographystyle{jfm}

\bibliography{bibliography}

\end{document}